\documentclass[twocolumn]{aastex63}

\hypersetup{linkcolor=red,citecolor=magenta,filecolor=cyan,urlcolor=blue}
\usepackage[T1]{fontenc}
\usepackage{ae,aecompl}

\usepackage{graphicx}	% Including figure files
\usepackage{amsmath}	% Advanced maths commands
\usepackage{amssymb}	% Extra maths symbols

\usepackage{xspace}
\usepackage{threeparttable}
\usepackage{color}

\usepackage{multirow}
\usepackage{enumitem}

\newcommand{\ergs}[1]{$\times 10^{#1}$ erg s$^{-1}$}
\newcommand{\oergs}[1]{$10^{#1}$ erg s$^{-1}$}

\newcommand{\ltsima}{$\buildrel < \over \sim$}
\newcommand{\lsim}{\lower.5ex\hbox{\ltsima}}
\newcommand{\gtsima}{$\buildrel > \over \sim$}
\newcommand{\gsim}{\lower.5ex\hbox{\gtsima}}

\newcommand{\swift}{{\it Swift}\xspace}

\newcommand{\xmm}{{\it XMM-Newton}\xspace}

\newcommand{\cxo}{\hbox{Chandra}\xspace}

\newcommand{\ulx}{\hbox{M51 ULX-7}\xspace}

\received{\today}
\submitjournal{ApJ}

\shorttitle{M51 ULX-7: off states and X-ray dips}
\shortauthors{Vasilopoulos et al.}
\graphicspath{{./}{figures/}{plots/}}
\begin{document}

% \title{M51 ULX-7: discovery of an irregular super-orbital low flux state and X-ray dips associated with the binary orbital period}
% \title{Evidence of an irregular super-orbital low flux state and X-ray dips associated with the binary orbital period it the ultraluminous X-ray pulsar M51 ULX-7}
% \title{Study of X-ray variability of M51 ULX-7: evidence for irregular super-orbital low flux state and X-ray dips associated with the binary orbital period}

% \title{Study of X-ray variability of M51 ULX-7: evidence for propeller transition and X-ray dips on orbital periods}

\title{Chandra probes the X-ray variability of M51 ULX-7: evidence of propeller transition and X-ray dips on orbital periods}

\correspondingauthor{Georgios Vasilopoulos}
\email{georgios.vasilopoulos@yale.edu}% $^{a}$Plaskett Fellow}

\author[0000-0003-3902-3915]{Georgios Vasilopoulos}
\affiliation{Department of Astronomy, Yale University, PO Box 208101, New Haven, CT 06520-8101, USA}

\author[0000-0002-1989-7984]{Filippos Koliopanos}
\affiliation{Universit{\'e} de Toulouse; UPS-OMP; IRAP, 31058 Toulouse, France}

\author[0000-0002-0107-5237]{Frank Haberl}
\affiliation{Max-Planck-Institut f\"ur extraterrestrische Physik,Giessenbachstra{\ss}e, 85748 Garching, Germany}

\author[0000-0003-0660-9776]{Helena Treiber}
\affiliation{Department of Astronomy, Yale University, PO Box 208101, New Haven, CT 06520-8101, USA}
\affiliation{Department of Physics and Astronomy, Amherst College, C025 New Science Center, 25 East Dr., Amherst, MA 01002-5000, USA}

\author{Murray Brightman}
\affiliation{Cahill Center for Astronomy and Astrophysics, California Institute of Technology, 1216 East California Boulevard, Pasadena,CA 91125, USA}

\author[0000-0001-5857-5622]{Hannah P. Earnshaw}
\affiliation{Cahill Center for Astronomy and Astrophysics, California Institute of Technology, 1216 East California Boulevard, Pasadena,CA 91125, USA}

\author{Andrés Gúrpide}
\affiliation{Universit{\'e} de Toulouse; UPS-OMP; IRAP, 31058 Toulouse, France}

% \author{\gv{Add your info here, preliminary order}}



\begin{abstract}
We report on the temporal properties of the ULX pulsar M51\,ULX-7 inferred from the analysis of the 2018-2020 \swift/XRT monitoring data and archival \cxo data obtained over a period of 33 days in 2012. 
We find an extended low flux state, which might be indicative of propeller transition, lending further support to the interpretation that the NS is rotating near equilibrium.
Alternatively, this off state could be related to a variable super-orbital period.
Moreover, we report the discovery of periodic dips in the X-ray light curve that are associated with the binary orbital period. 
The presence of the dips implies a configuration where the orbital plane of the binary is closer to an edge on orientation, and thus demonstrates that favorable geometries are not necessary in order to observe ULX pulsars.
These characteristics are similar to those seen in prototypical X-ray pulsars like Her X-1 and SMC X-1 or other ULX pulsars like NGC 5907 ULX1.
\end{abstract}


\section{Introduction}\label{sec:intro}

Ultra luminous X-ray (ULX) sources \citep{2017ARA&A..55..303K} are off-nuclear extra-galactic X-ray binary systems with an apparent isotropic luminosity that exceeds the Eddington limit for an accretion powered, stellar mass black hole (i.e.~L$_X{>}$\oergs{39}). 
Given their high luminosity ULXs were thought to host the elusive intermediate-mass black holes.
Remarkably, within the last years there has been undisputed evidence that at least a few of these systems are powered by accreting highly magnetized neutron stars (NS); these are known as ULX pulsars \citep[ULXPs, ][]{bachetti_ultraluminous_2014,furst_discovery_2016,israel_accreting_2017,israel_discovery_2017, carpano_discovery_2018,2020ApJ...895...60R}.
This discovery is consistent with theoretical predictions \citep[e.g.][]{1976MNRAS.175..395B,2015MNRAS.447.1847M} that argue that NSs can break the barrier set by the Eddington limit ($L_{\rm Edd}\sim{1.4\times10^{38}M/M_{\odot}}$ erg/s) for strong magnetic fields ($B$).
Moreover, an increasing number of authors have put forward the proposition that a major fraction of ULXs are powered by NSs rather than black holes \citep[e.g.][]{2017MNRAS.468L..59K,2017A&A...608A..47K,2018ApJ...856..128W}.

\begin{figure*}
    \vspace{-0.4cm}
	\resizebox{1.0\hsize}{!}{
	\includegraphics*[width=1.99\columnwidth]{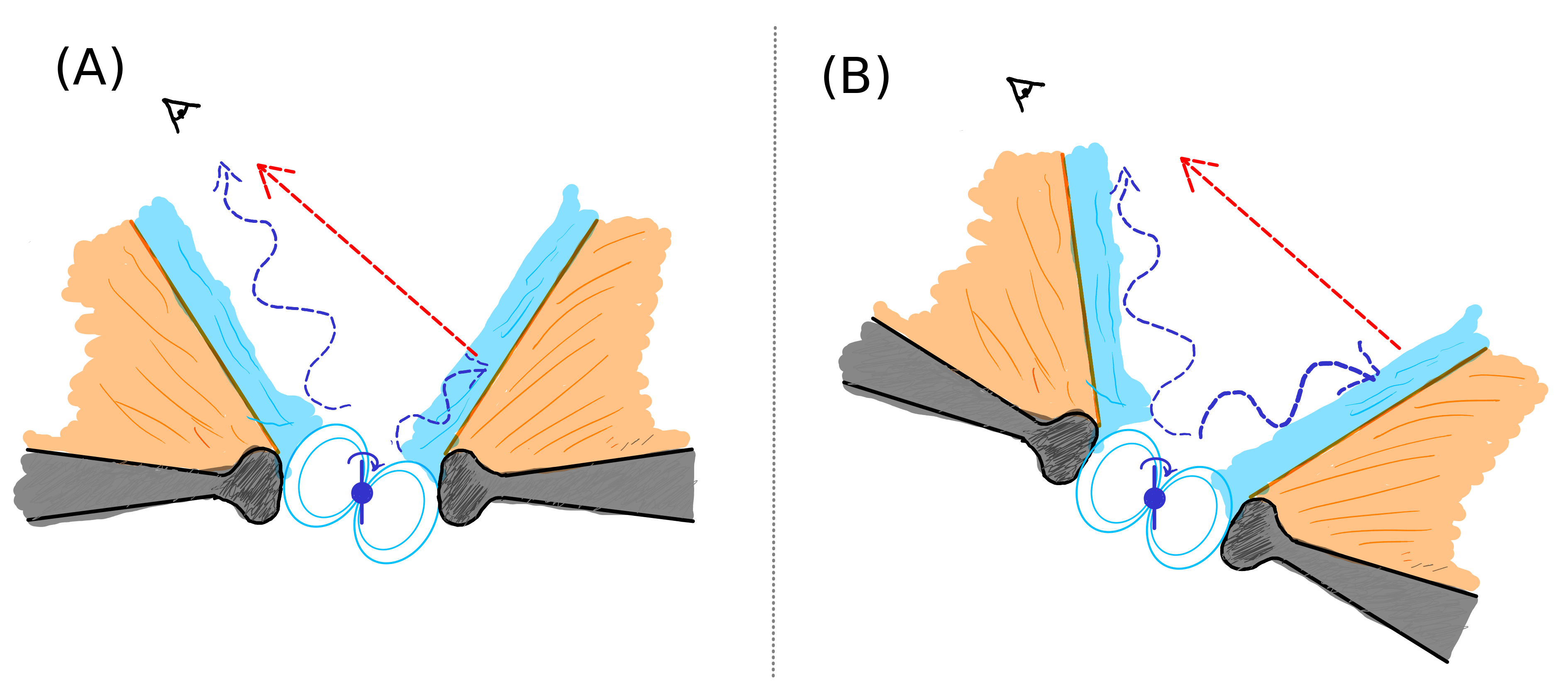}}	
	\vspace{-0.3cm}
    \caption{\label{fig:1} Schematic of outflows from an accretion disk during super-Eddington accretion. Outflows can start from the disk (i.e.~orange shade), or even inside the magnetosphere (i.e.~cyan shade) given the super-Eddington luminosity.
    The walls of the outflow create a funnel and an observer can see the central source if it is in a favorable orientation (i.e.~left panel). The pulsed emission is shown with blue dotted lines, while reprocessed emission is shown with red dotted lines. Given the extent of the funnel, coherent pulsations are diluted, while the spectral shape of the emission should also alter.
    If the disk precesses, the funnel also follows the same motion, and thus the observer sees a super-orbital modulation. If the observer's line of sight is obscured by the funnel walls, then only non-pulsating emission from the funnel walls should be visible (i.e.~right panel).}
\end{figure*}

For a standard accretion disk \citep{1973A&A....24..337S}, as the accretion rate reaches the Eddington limit, the radiation pressure dominates the inner part of the accretion disk, causing a large fraction of the accreted material to be lost through outflows \citep{2007MNRAS.377.1187P}.
Material is expelled inside  the spherization radius $R_{\rm sph}$, and the outflow is not spherical but forms a funnel-like structure (see Fig.~\ref{fig:1}).
In the context of ULXPs, the disk is truncated at the magnetospheric radius $R_{\rm M}$. For high $B$ values, truncation can therefore occur outside $R_{\rm sph}$.
However, it has been proposed that outflows could form in a similar manner inside $R_{\rm M}$, as material is accreted onto the NS via the magnetic field lines \citep[][]{2017MNRAS.468L..59K}.

% \begin{figure*}
%     \vspace{-0.4cm}
% 	\resizebox{1.0\hsize}{!}{
% 	\includegraphics*[width=1.99\columnwidth]{ULX_funnel_pre.png}}	
% 	\vspace{-0.3cm}
%     \caption{\label{fig:1} Schematic of outflows from an accretion disk during super-Eddington accretion. Outflows can start from the disk (i.e.~orange shade), or even inside the magnetosphere (i.e.~cyan shade) given the super-Eddington luminosity.
%     The walls of the outflow create a funnel and an observer can see the central source if it is in a favorable orientation (i.e.~left panel). The pulsed emission is shown with blue dotted lines, while reprocessed emission is shown with red dotted lines. Given the extent of the funnel, coherent pulsations are diluted, while the spectral shape of the emission should also alter.
%     If the disk precesses, the funnel also follows the same motion, and thus the observer sees a super-orbital modulation. If the observer's line of sight is obscured by the funnel walls, then only non-pulsating emission from the funnel walls should be visible (i.e.~right panel).}
% \end{figure*}

Super-orbital modulation of ULXs is possible through precession of the funnel \citep{2017MNRAS.466.2236D}.
For three known ULXPs (M51\,ULX-7, M8\,2 X-2 and NGC\,5907 ULX1) super-orbital periodicities (40 d, 60 d and 78 d, respectively) are evident in their X-ray light curves \citep[see][and references therein]{2020ApJ...895..127B}. 
The observed flux ($F_{\rm X}$) during a super-orbital cycle can vary by a factor of 100, but there has been no concrete evidence for  spectral changes indicating accretor-to-propeller transitions \citep{1975A&A....39..185I}.
These transitions occur when the inner disc radius (i.e.~the magnetosphere of the NS) becomes larger than the corotation radius of the NS, the centrifugal drag causes material to be propelled away instead of accreted. 
Alternatively, the changes in $F_{\rm X}$ can be due to obscuration by a precessing accretion disk and a funnel formed by optically thick outflows  
\citep{2018MNRAS.475..154M}.
A firm confirmation of this scenario has been shown in the case of NGC 300\,ULX1, where a stable spin-up rate has been maintained during epochs of variable $F_{\rm X}$ \citep{2019MNRAS.488.5225V}.
However, the engine behind precession is still unclear; it could be the tidal force from the massive companion star, the interaction with the magnetosphere of the NS, the irradiation of the warped disc, or even the NS free precession \citep[see][and references within]{2020MNRAS.491.4949V}. Constraints in theoretical models can only be derived by monitoring ULXPs and their super-orbital periodicity, the study of the stability of this periodicity, and the occurrence of on and off states.
One of the unanswered fundamental questions about ULXPs is if their X-ray emission is beamed towards the observer, and what is the beaming factor, i.e.~apparent luminosity is larger than the true on $L_{\rm app}{=}bL$.
For a narrow funnel (see Fig.~\ref{fig:1}) beaming could be very strong, while for wide funnels the beaming factor could be smaller than 2. This means that the bolometric X-ray emission of the source is only overestimated by a small factor. Given the lack of physically self-consistent spectral models and the degeneracy of phenomenological ones \citep[e.g.][]{2017A&A...608A..47K,2019A&A...621A.118K}, we should perhaps look at temporal properties for observational constraints on $b$.

\begin{figure*}
	\resizebox{1.0\hsize}{!}{
	\includegraphics*[width=1.99\columnwidth]{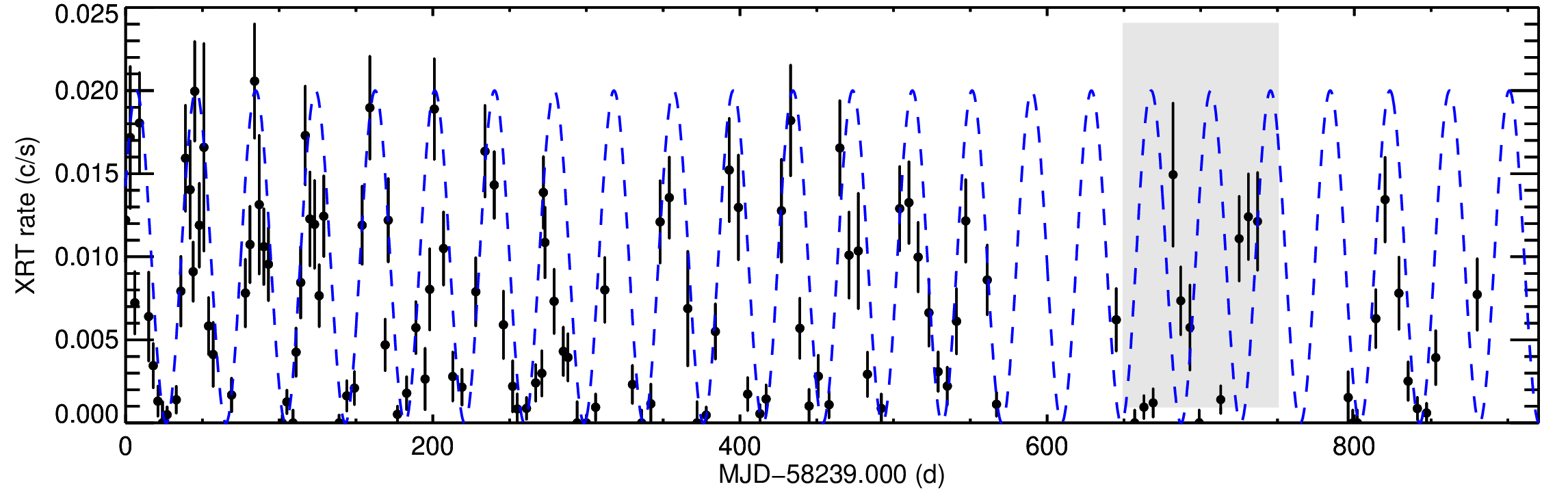}}	
	\vspace{-0.3cm}
    \caption{\label{fig:22} X-ray light curve of \ulx based on the 2018-2020 \swift/XRT monitoring of the region. Points below 0.005 c/s may be considered as non-detection or upper limits as they correspond to less than $\sim$5-10 total counts. A sinusoidal curve with a period of 38.86 d is plotted to guide the eye. 
    }
\end{figure*}

\begin{figure}
	\resizebox{1.0\hsize}{!}{
	\includegraphics*[width=1.99\columnwidth]{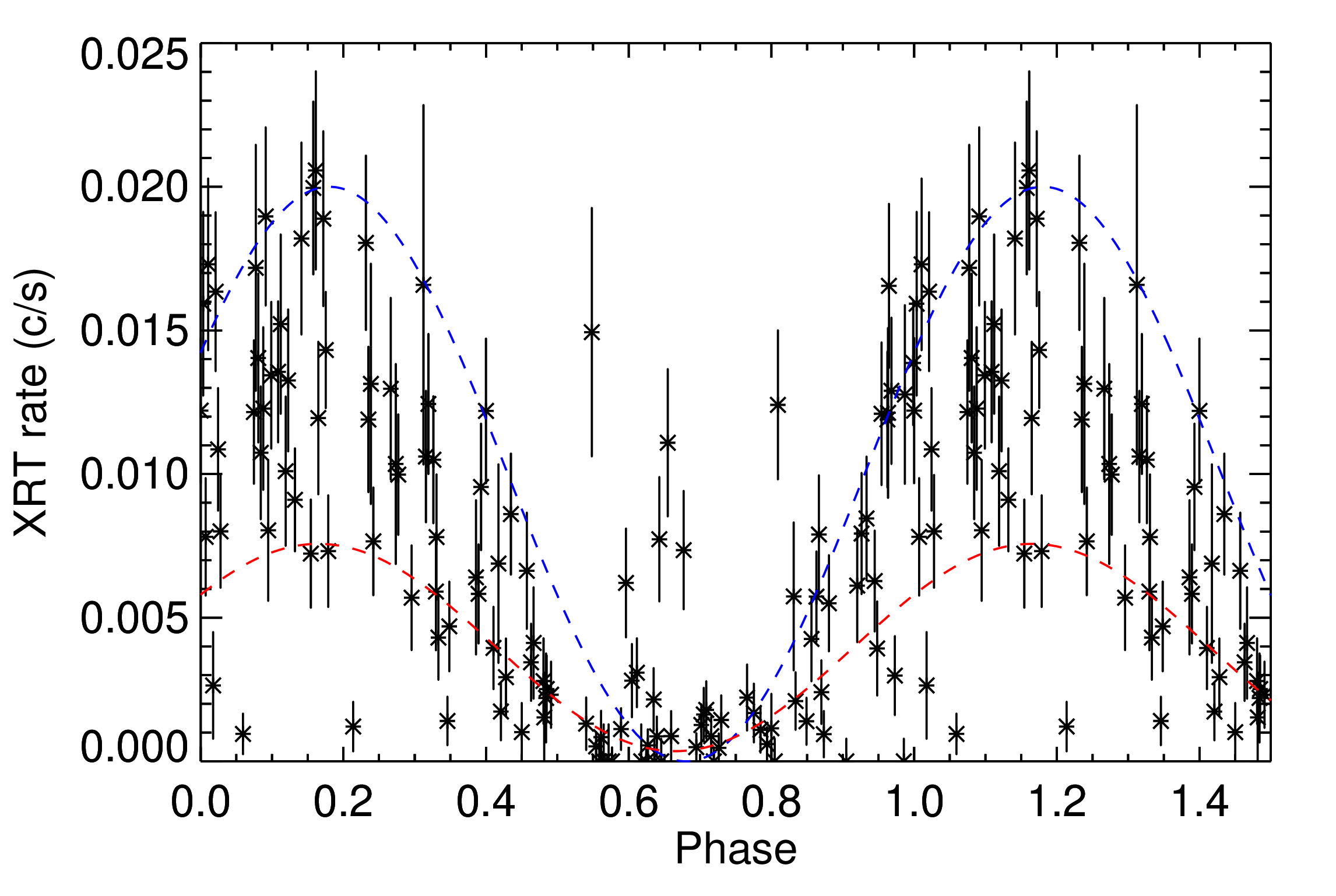}}	
	\vspace{-0.3cm}
    \caption{\label{fig:33} X-ray lightcurve of Fig. \ref{fig:22} folded for the super-orbital period of 38.9 days. Blue dotted curve is same as Fig. \ref{fig:22}. Red line is scaled to match the lower flux points through the super-orbital cycle.
    Most points follow the super-orbital trend, while few outliers are mainly from observations between MJD 58900-59000\,d (see shaded area in Fig. \ref{fig:22}). 
    }
\end{figure}

\citet{Gurpide2021} have processed archival data of 17 ULXs to investigate their long-term X-ray spectral evolution. Motivated by their work and using the products of their analysis, we studied the variability of \ulx, the only ULXP with an orbit that can be continuously monitored by X-ray observatories \citep{2020MNRAS.495L.139T}. 
\ulx is a ULXP \citep{2020ApJ...895...60R} hosting a NS rotating with a spin period of $\sim$2.8\,s.
The binary period is 1.9969 d, while \swift/XRT monitoring revealed the presence of a super-orbital modulation with a period of $\sim$38-39~d \citep{2020ApJ...895..127B,2020MNRAS.491.4949V}. 
Here we report on the discovery of an irregular off-state within the super-orbital cycle of \ulx and the discovery of periodic dips in the X-ray light curve computed from archival \cxo data.
The detection of eclipses offers new insights onto the geometrical configuration of ULXPs, and could provide an independent constraint on the beaming factor of \ulx.

\section{Data Analysis \& Results}\label{sec:data}

\subsection{Probing the super-orbital modulation}
For this study we used X-ray data from the M51 monitoring \citep{2020ApJ...895..127B} by the 
Neil Gehrels \swift\ Observatory \citep[][]{2004ApJ...611.1005G} X-ray Telescope \citep[XRT,][]{2005SSRv..120..165B}. 
\swift/XRT data were analysed following the pipeline developed by \citet{2020ApJ...895..127B}. 
To estimate an updated super-orbital period we computed the Lomb-Scargle (LS) periodogram \citep{1982ApJ...263..835S} for the 2018-2020 data shown in Fig. \ref{fig:22}. A period of $\sim$38.86\,d was found. 
% while the super-orbital maximum flux is expected at MJD 58246.017$\pm$N$\times$38.86\,d, where N is an integer.
% while the super-orbital maximum flux is expected at MJD 58246.017$\pm$N$\times$38.86\,d, where N is an integer.
% 
A systematic uncertainty on the derivation of the super-orbital period has to do with the treatment of marginal detections by \swift/XRT. By following \citet{2020MNRAS.491.4949V} many low flux points are consistent with upper limits. By ignoring all lower flux points with rates smaller than 0.005 c/s, the LS periodogram yields a periodicity of 38.94 d. The difference between the two methods may be considered as an estimate of the uncertainty of the super-orbital period. Thus the super-orbital period should be $\sim$38.9~d, while the super-orbital maximum flux is expected at MJD 58246.017$\pm$N$\times$38.9\,d, where N is an integer.
% 
% To estimate the uncertainty we followed a bootstrapping method and repeated the LS calculations for 1000 samples.
% The uncertainty of the computed super-orbital period is about $\sim$0.08 d. 
% 
Compared to the previous studies of \ulx \citep[see,][]{2020MNRAS.491.4949V,2020ApJ...895..127B}, a super-orbital periodicity is evident in the updated data set. However, for a 100 day interval around MJD 58900-59000\,d the observed modulation appears to fall out of phase compared to the general trend (see shaded area in Fig. \ref{fig:22}), before returning to the normal phase at a later time (i.e. around MJD 59040\,d).

\begin{figure*}
	\resizebox{1.0\hsize}{!}{
	\includegraphics*[width=1.99\columnwidth]{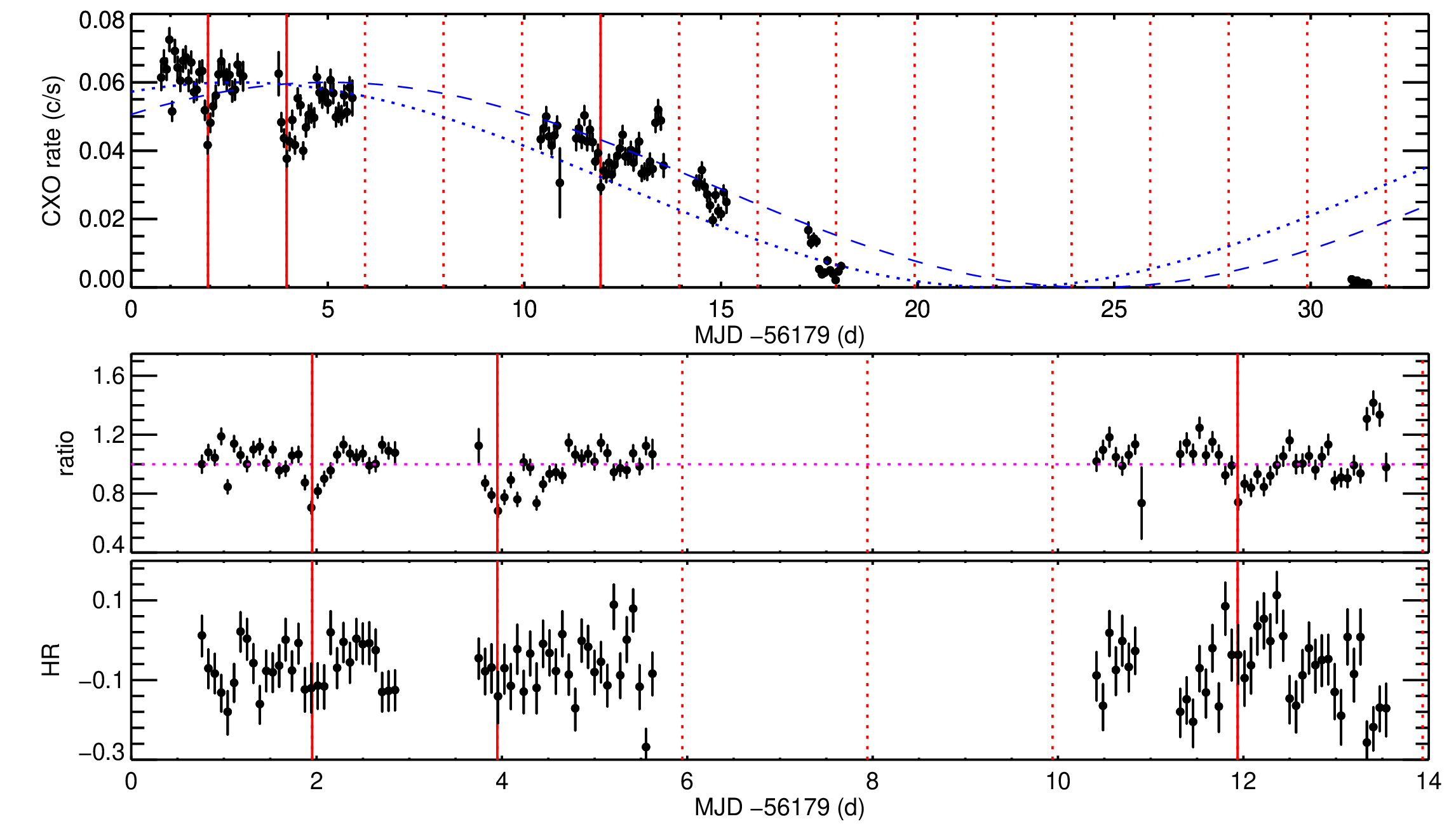}}	
	\vspace{-0.3cm}
    \caption{\label{fig:2} \emph{Upper panel:} X-ray light curve (0.3-8.0 keV) of \ulx based on \cxo data obtained in 2012. Events are binned every 6000\,s. The vertical dotted lines are phased with the binary orbit of 1.9969 d. There is an indication of periodic flux drops occurring at the same orbital phase. The blue dashed line marks the 38.9 d super-orbital modulation similarly to Fig. \ref{fig:22} and Fig. \ref{fig:33}. 
    % The dotted line marks the uncertainty of the extrapolated solution, based on the super-orbital period uncertainty, i.e. 38.94 d.
    The dotted line marks the extrapolated solution for a 38.94 d period (see text for details).
    \emph{Middle panel:} Ratio between \cxo data and a linear model fit to the first 15 days of the \cxo monitoring.
    % \emph{Middle panel:} Ratio between \cxo data and the extrapolated super-orbital modulation for the first 16 days of the \cxo monitoring.
    \emph{Lower panel:} Spectral hardness evolution estimated by HRs. There is only a marginal indication of spectral softening in the first dip.
    }
\end{figure*}

M51 was observed by \cxo 16 times.
Most of these observations have short exposure times (e.g.~$\sim$20\,ks) and thus only contribute to the study of the variability of \ulx on timescales of a few hours \citep[i.e.~2 hour long dips reported by][]{2002ApJ...581L..93L}.
M51 was observed by \cxo between 2012 September 9 and October 10 (PI: Kuntz) with a few long visits (i.e.~$>$100 ks). \citet{2016MNRAS.456.3840E} analysed the 2012 data and reported on the properties of \ulx, but the X-ray light curves were not presented in their work. In Table \ref{tab:cxo} we present the \cxo  observations that we used in our analysis, these are C6-C12 as defined in Table 1 of \citet{2016MNRAS.456.3840E}. 
Data reduction was performed with the {\tt CIAO} software \citep{2006SPIE.6270E..1VF}. For creation lightcurves, the source (background) events were extracted from circular regions with 3 (20) arcsec radius \citep[following][]{2016MNRAS.456.3840E}. We used standard \texttt{FTOOLS} scripts to perform event selection and create light curves. For the light curve we used a 6000\,s binning to compromise between getting acceptable statistics and sufficient timing resolution (see Fig.~\ref{fig:2}). 
To investigate spectral changes we estimated the spectral Hardness Ratio (HR) from each 6000\,s interval.
We define HR as the ratio of the difference over the sum of the number of counts in two subsequent energy bands: $\rm{HR}{=}(\rm{R}_{\rm{i+1}}-\rm{R}_{\rm{i}})/(\rm{R}_{\rm{i+1}}+\rm{R}_{\rm{i}})$,   
where $\rm{R}_{\rm{i}}$ is the background-subtracted count rate in a specific energy band (i.e.~0.3-1.5 keV and 1.5-8.0 keV).
HRs were computed with a Bayesian estimator tool \citep[\texttt{BEHR};][]{2006ApJ...652..610P}.

In Fig.~\ref{fig:2} we plot the X-ray light curve obtained from the \cxo observations in September 2012.
The modulation looks consistent with the reported 38.9 d super-orbital modulation of \ulx \citep{2020MNRAS.491.4949V}. To test this we extrapolated the super-orbital solution derived by fitting a sinusoidal function to the \swift/XRT monitoring data, as shown with a blue line in Fig.~\ref{fig:22}.  
We scaled the function by a factor of 3, which is the ratio of the ACIS-S to XRT spectral response given the spectral properties reported by \citet{2016MNRAS.456.3840E}, i.e. an absorbed power-law with $\Gamma{=}1.5$ and $N_{H}$=1.5$\times$10$^{21}$cm$^{-2}$. 
The result is shown in Fig.~\ref{fig:2} by the dashed blue line. For clarity, we also plot a sinusoidal function computed for the upper limit of the super-orbital period, i.e. 38.94 d.
The agreement with the \cxo data is good for the first part of the light curve. 
However, during the final \cxo observation (obsid: 15553) the flux remained at a low level.
To further investigate the drop in flux within the last observation we derived accurate count rates for each \cxo pointing.
We performed source detection using the {\tt CIAO} {\tt wavdetect} tool that implements a wavelet analysis on the X-ray images. The resulted count rates are given in Table \ref{tab:cxo}. The drop in flux seen around MJD 56210 (obsid: 15553) can be compared with observations taken about 15 days earlier (obsid: 13815), when the flux should have been similar according to the super-orbital modulation. Thus  we find that on MJD 56210 the flux of \ulx is $\sim$80 times lower than expected.

\begin{table}
\caption{\cxo Observing log\label{tab:cxo}}
\begin{threeparttable}[b]
\begin{tabular*}{\columnwidth}[t]{p{0.15\columnwidth}p{0.2\columnwidth}p{0.20\columnwidth}p{0.20\columnwidth}}
\hline
Obsid$^{(a)}$ & Date & Exposure &  Rate$^{(b)}$\\
 &  & ks & 10$^{-2}$ c/s\\
\hline
\hline
13812  & 2012-09-12  &  159 & 5.07$\pm$0.06\\ 
13813  & 2012-09-09  & 181  &  6.00$\pm$0.06\\ 
13814  & 2012-09-22  &  192 &  3.88$\pm$0.05\\ 
13815  &  2012-09-23 &  68 &  2.58$\pm$0.06\\ 
13816  & 2012-09-26  &  74 &  0.65$\pm$0.03\\ 
15496  & 2012-09-19  & 42  &   4.44$\pm$0.10\\ 
15553  & 2012-10-10  & 38 & 0.032$\pm$0.010\\
 \hline\noalign{\smallskip}  
\end{tabular*}
\tnote{(a)} \footnotesize{All data were obtained by ACIS-S camera.}
% \tnote{(b)} \footnotesize{Total counts within a 3\arcsec region and the 0.3-8.0 keV band, the average background count-rate for the source extraction region is $\sim$0.4-0.7 counts/ks.}
\tnote{(b)} \footnotesize{Net count rates  (0.3--8 keV band) derived from the {\tt wavdetect} tool.}
\end{threeparttable}
\end{table}

\begin{figure*}
	\resizebox{1.00\hsize}{!}{
	\includegraphics*[width=1.99\columnwidth]{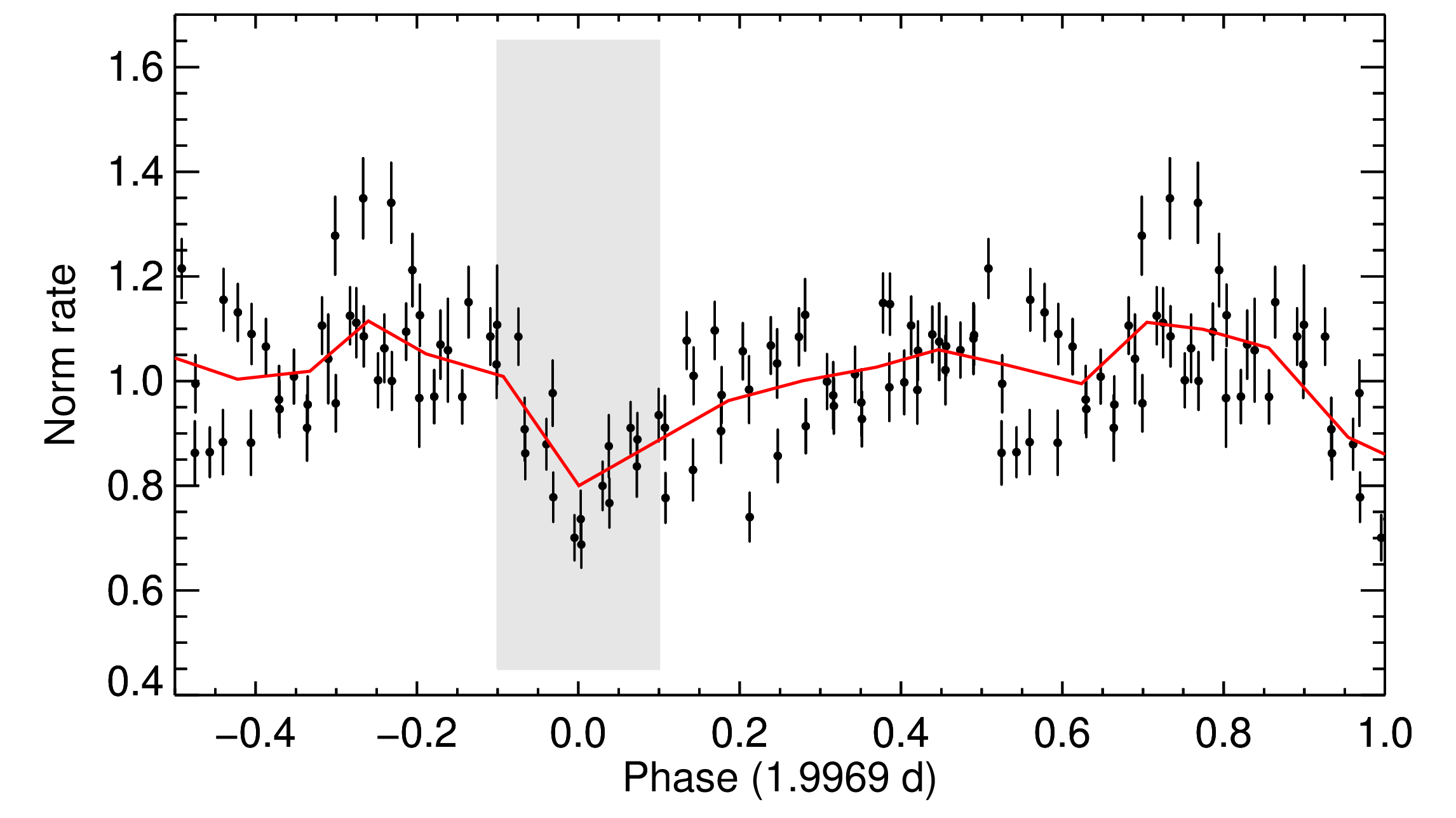}
	\includegraphics*[width=1.99\columnwidth]{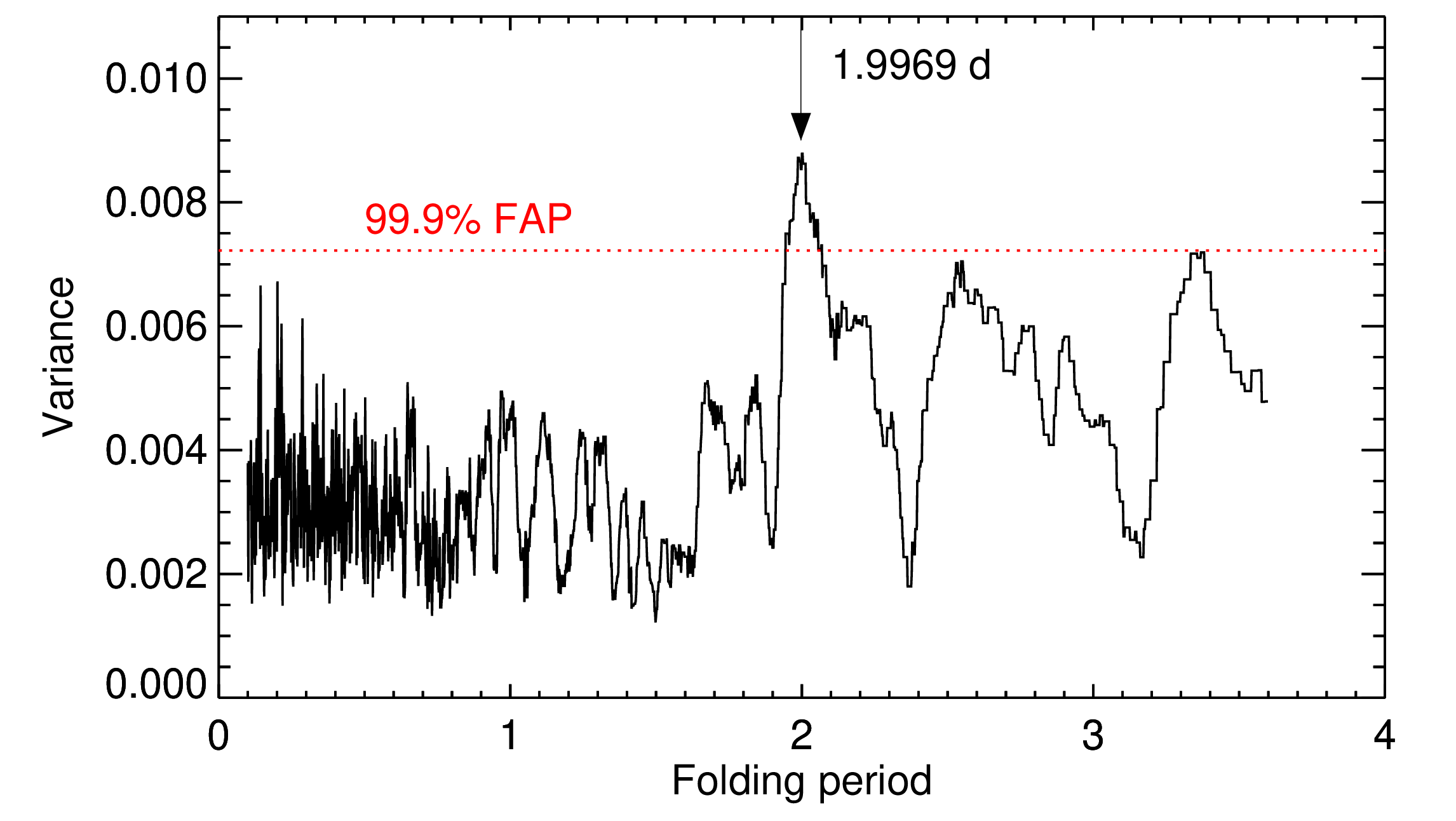}
	}	
	\vspace{-0.5cm}
    \caption{\label{fig:3} \emph{Left:} \cxo X-ray ratio light curve of \ulx folded for the orbital period of the binary. We only used data from obsids 13812-4. Normalised rates were calculated in reference to a linear model fitted to the data. The shaded region marks the approximate duration of the dips.
    \emph{Right:} Result of epoch-folding method. The maximum variance is found for exactly the orbital period of the system.
    }
\end{figure*}

\subsection{Detection of periodic X-ray dips}

The \cxo light curve shows fast variability on timescales of a few 1000 s, which is consistent with other studies of the system \citep{2016MNRAS.456.3840E,2002ApJ...581L..93L}. In particular \citet{2002ApJ...581L..93L} claimed the presence of a 2 hour periodicity in early \cxo data, that however only had 20 ks exposure time. Looking at isolated chunks of data it is easy to find such trends, like dips and flares, that nevertheless could be random in nature. However, knowing that the system has a 1.9969\,d orbital period we can guide the eye to identify any patterns. By doing so, we identify that the three strongest dips appear to be periodic. These are marked with vertical red lines in Fig.~\ref{fig:2}. These three dips are not defined by a single point, but have a structure that resembles a trough. Moreover, the first of these appear to be the strongest, but this could also be related to the better statistics obtained in the first \cxo visit, which occurred during the maximum of the super-orbital phase. 
For visualization purposes, we detrended the lightcurve by fitting a linear regretion model to the data, this is shown in middle panel of Fig.~\ref{fig:2}.
Finally, we have folded the data of the first 15 days with the orbital period (see Fig.~\ref{fig:3}). The drop in flux during the dips is of the order of 20-30\%, but given the intrinsic difficulties in determining a baseline flux, this should be considered as an upper limit. 

To investigate the statistical significance of the periodic dips we used various tests based on epoch folding \citep{1990MNRAS.244...93D}. 
The data were folded for a series of test periods, and the resulting profiles were then tested against constancy using a $\chi^2$ test, or a maximization of variance. 
We used the normalised ratio light curve produced from the three observations where the dips are distinguished (obsids: 13812-4). We folded the ratio light curve for test periods between 0.5 and 3.5 days. Then we binned the folded profile to obtain 20 average measurements (similar to the smoothed profile in Fig.~\ref{fig:3}) and we calculated the variance of these average values. Finally, we repeated the procedure for 10000 simulated data-sets to estimate the false alarm probability.
The result is plotted in the right panel of Fig.~\ref{fig:3}.

Given that the statistical significance of the X-ray dips has been established we should address whether these can be associated with a specific phase of the orbital period.
The orbital ephemeris of the system was determined using \xmm data obtained between MJD 58251 and 58281\,d \citep{2020ApJ...895...60R}.
Given that the \cxo data were obtained about 1100 orbits before that, the ephemeris cannot be extrapolated with enough accuracy to actually compare the predicted transitions of the optical star in front of the NS (i.e. time of ascending nodes: $T_{\rm asc}$) with the X-ray dips. For such comparison we used the the \swift/XRT data from the 2018-2020 monitoring of the system (MJD 58000-59125\,d). Given that individual observations can span up to one day, we performed source detection to individual snapshots and folded the resulting light curve with the orbital period. 
We found no evidence of X-ray dips in the \swift/XRT orbit-folded light curve. Nevertheless, this should be expected due to the low effective area of XRT and the short exposures that resulted in high uncertainties for individual detections. Specifically for the 170 snapshots where the source was detected, the uncertainties were of the order of 38$\pm$15\%, i.e. similar or larger than the expected drop in flux during the dips seen in \cxo data (i.e.~15-30\%).

\section{Discussion}

%  ---> Discussion
Following the discovery of pulsations originating from the NS in \ulx, it has been shown that in order to model the spectral and temporal properties of the system self-consistently, the NS should have a very strong magnetic field and should be in the fast rotator regime \citep{2020MNRAS.491.4949V}. \citet{2020MNRAS.491.4949V} also argued that the $\sim$39\,d super-orbital modulation of the ULXP could be triggered by (or related to) free precession of the NS, which surprisingly requires a NS magnetic field of 3-4$\times$10$^{13}$ G, in quantitative agreement with the spin equilibrium predictions. As a consequence, strong outflows are not expected from the accretion disk (since $R_{\rm M}$>$R_{\rm sph}$). Thus the opening funnel of any outflow should be large, and as a result we do not expect strong beaming by the system and accretion onto the NS is indeed 10-30 times above the Eddington limit \citep{2020MNRAS.491.4949V}. 
In the following paragraphs, we will discuss how the newly reported patterns of X-ray variability can be interpreted and address some of the open questions for \ulx and ULXPs, like investigating beaming and the engine behind super-orbital modulation. 

\subsection{Propeller transition or an unstable super-orbital clock}
\label{sec:prop}

In terms of super-orbital variability, the agreement between the modulation seen in the \cxo data and the phase of the expected maximum of the super-orbital period computed by the \swift/XRT monitoring data is remarkable. Especially considering that the \swift and \cxo data are separated by about 50 super-orbital cycles.
Thus, the drops may be related to a transition to the propeller state, as it is speculated for similar drops in flux seen in NGC 5907 ULX1
\citep{2017ApJ...834...77F}.
This drop in flux is in agreement with the behavior seen by \swift/XRT around MJD 55715\,d \citep[see][]{2020MNRAS.491.4949V}, where during a 70 day monitoring \ulx was not detected in the final 20 days of the monitoring, when the rise of its flux was expected according to the super-orbital period. 
However, in both epochs, there are no additional monitoring data to determine the duration of the off state, or investigate if the super-orbital period was different than the one determined by the 2018-2020 monitoring. 
A similar ``off-state'' is seen in the \swift/XRT data around MJD 58900\,d (see Figs. \ref{fig:22}, \ref{fig:33}), where the flux dropped near zero for 3 consecutive visits. In that case although the super-orbital cycle seemed to be disturbed the super-orbital period returned to its normal beating pace after a few cycles. 

A stable super-orbital period would also be in agreement with the requirements of the NS free precession mechanism, as the ratio of super-orbital and NS spin period should be proportional to the NS $B$ field.
Since a transition to the propeller regime can occur with minimal change in mass accretion if the accretion disk is truncated near the NS corotation radius, this drop in flux lends further support to the findings of \citet{2020MNRAS.491.4949V}, who proposed that the NS has a magnetic field $\sim3-7\times10^{13}$~G and is rotating near its equilibrium period.
Nevertheless, a variable super-orbital period cannot be excluded for \ulx. If a variable period is confirmed with future monitoring data this would reveal a similar observational behavior to HMXB pulsars like SMC X-1 \citep{2007ApJ...670..624T}.

% A small beaming factor can also be addressed in the context of propeller transition (see \S\ref{sec:prop}).  
Assuming that the off-state is associated with propeller transition, we can derive the NS magnetic field $B$ that would be required for the transition to occur.
Following \citet{2018A&A...610A..46C} we find that:
\begin{equation}
    B=10^{12} \left(\frac{L_{X,min}}{2\times10^{38} erg/s}(P/1 s)^{7/3}{\xi}^{-7/2}\right)^{1/2} G,
    \label{eq:prop}
\end{equation}
where $\xi$ is a normalization factor with typical value of 0.5 \citep[however see case of ULXs][where $\xi$ can be higher]{2019A&A...626A..18C}, and $L_{\rm X,min}$ is the minimum luminosity before the propeller transition. 
Given that the super-orbital modulation is due to precession (see Fig. \ref{fig:1}), its true $L_{\rm X}$ is similar to the maximum X-ray luminosity within the super-orbital cycle. 
For \ulx this is of the order of 7\ergs{39} in the 0.3-10.0 keV band \citep{2020ApJ...895...60R,Gurpide2021}, however for ULXPs the bolometric luminosity could be of a factor of 2 higher \citep{2017A&A...608A..47K,Gurpide2021}. 
Moreover, it has already been established that the observed $L_{\rm X}$ is only boosted by small beaming \citep{2020MNRAS.491.4949V}, of the order of 2 or less (see also \S\ref{sec:beam}).
% 
% Finally, given that \ulx is near equilibrium propeller transition can occur for small changes (factor $<2$) in the intrinsic $L_{\rm X}$.
Since many of the above uncertainties cancel out, we can adopt a value of $L_{\rm X,min}\sim7$\ergs{39} for propeller transition (but we caution the reader for the above mentioned uncertainties), thus based on equation \ref{eq:prop} we find $B\sim{4-9}{\times}10^{13}$ G.
We note that, depending on the torque model used the propeller transition can occur very close (factor $<2$)  to the $L_{\rm X}$ that probes spin equilibrium for a given magnetic field \citep[see comparisons][]{2016ApJ...822...33P,2018A&A...620L..12V}.
Thus, equation \ref{eq:prop} yields a similar magnetic field estimate to the value derived by just assuming the NS is rotating near equilibrium \citep[see discussions for \ulx and M82\,X-2][]{2020MNRAS.491.4949V,2015MNRAS.448L..40E}.

So far we have connected the ``off-states'' of \ulx with the source transitioning from accretor to propeller regime. Thus, during the ``off-state'' the source has moved way down on the Luminosity gap for X-ray pulsars \citep[i.e., Corbet gap;][]{1996ApJ...457L..31C}. However, given the spin period of $\sim$2.8 s such transition should have resulted in a drop\footnote{Luminosity jump is propotional to the ratio of the dynamical energy at the NS surface over the corotation radius \citep{1996ApJ...457L..31C}, i.e. $\sim170\times$(P/1 s)$^{2/3}$} in $L_{\rm X}$ of the order of $\sim$330, and not just $\sim80$. This discrepancy may be explained if we assume that even at the propeller regime, there is still some residual accretion that can penetrate the magnetophseric barrier \citep{1993ApJ...402..593S}, a mechanism that has been supported by theory and simulations \citep[e.g.,][]{2012MNRAS.420..416D,2017ApJ...851L..34P,2018NewA...62...94R}.

An opposing view would be that during the ``off-state'', \ulx is still in the accretor regime. This limit may be used to put a lower limit to the propeller stage assuming that the source is then on the propeller line. By using equation \ref{eq:prop} we find $B{\sim}5\times10^{12}$~G. However, this low $B$ field value posses difficulties in explaining the very low NS spin-up rate ($\dot{P}_{\rm NS}$) observed during maximum X-ray luminosity \citep{2020ApJ...895...60R}. Given all the observational evidence, to account for this low $B$ value and the observed $\dot{P}_{\rm NS}$, we would need change our basic assumptions (for the accretion disk) in order to decrease the rate of angular momentum transfer. 
Inefficient angular momentum transfer may be achieved if the accretion disk is miss-aligned with the $NS$ rotation axis, or the inner disk velocity significantly deviates from the Keplerian approximation.

\subsection{M51\,ULX-7 as an ULXP analog of Her X-1}
\label{sec:her}

The study of ULXPs has revealed not only that some host strongly magnetized NSs, but also that this might be the norm for a large fraction of ULXs \citep{2017A&A...608A..47K,2017MNRAS.468L..59K}. 
Thus, it is natural to look for similarities (and also differences) between individual ULXPs and HMXBs, which host the majority of X-ray pulsars. The NS spin period as well as the orbital and super-orbital periods of \ulx (2.8\,s, 1.99\,d and $\sim$40\,d), are close to the values (1.24\,s, 1.7\,d and $\sim$35\,d) of the prototypical X-ray pulsar Her X-1 \citep{1972ApJ...174L.143T,1973NPhS..246...87K}. In regards to the super-orbital modulation, for both systems it has been proposed that NS free precession could play an important role \citep{1986ApJ...300L..63T,2009A&A...494.1025S,2020MNRAS.491.4949V}. 
Another characteristic feature of Her X-1 is the presence of eclipses that coincide with the orbital period of the binary.
Our study of \ulx found evidence of similar features that occur periodically and could help to further constrain the properties of the system.
In order to understand the nature of the X-ray dips in \ulx we can thus refer to the plethora of theoretical models proposed for Her X-1. 
In Her X-1 full eclipses occur when X-rays are obscured by the companion star, however in its X-ray light curve there are characteristic X-ray dips, commonly referred to as \emph{pre-eclipse} or \emph{anomalous dips}. The pre-eclipse dips (2-5 h long) occur before the eclipse and gradually march backward in phase within the super-orbital cycle.
Anomalous dips (1-2 h long) occur at the same orbital phase, while there is evidence of cold matter absorption, in contrast to the pre-eclipse dips \citep{1995A&A...297..747R}. Theoretical explanations of these dips include dynamical, hydrodynamic and radiative interactions between the accretion stream, the warped accretion disk and the companion star \citep[e.g.][]{1996A&A...307...95S,1999A&A...348..917S}.  If the stream falls into the warped disk with an angle, the formation of a cold clumpy spray is possible, that in turn will cover the central source once per orbit \citep{1996A&A...307...95S}. In this scenario, a turbulent thickening of the warped disk is also possible. 

A different cause for the X-ray dips is obscuration by the stellar wind of the companion. 
For X-ray binaries it is important to take into account the strong X-ray illumination of the companion star that can affect the geometry of the stellar winds.
It has been shown that X-ray illumination can cause the formation of a so-called ``shadow wind'' by the companion in luminous HMXBs \citep{1994ApJ...435..756B}. By performing 2D hydrodynamic simulations, \citet{1994ApJ...435..756B} found that for high X-ray luminosities the gas that resides on the stellar surface exposed to the X-ray source will be highly photoionized, and thus halt the formation of a radiative driven wind from that side \citep[see][for an application to 4U\,1700-37]{1989ApJ...343..409H}. Given that stellar wind can still escape from the other side of the star, enhanced column density is still possible at favorable orientations and orbital phases. In fact, the shadow wind model has been offered as a possible mechanism to explain periodic dips seen during a super-Eddington outburst of SMC X-2 \citep{2016ApJ...828...74L}. 
For systems like Her X-1, the irradiation of the companion star can also depend on the binary orbital phase \citep{1999A&A...348..917S}. In this scenario, the (phase dependent) shadowing of the companion by the disk leads to the formation of flows coming out from the orbital plane and could in principle shadow the central source once per orbit. 

For \ulx it is possible to imagine a similar scenario, where either (or all) of the above mechanisms can provide the necessary conditions to form the X-ray dips under a favorable orientation. Nevertheless, the lack of significant spectral change during the dips (see HRs in Fig. \ref{fig:2}), could be an indication of obscuration from a fully ionised material rather than cold matter. 

% \subsection{Constraints on the beaming}
\subsection{Constraints on orbit inclination and ULX beaming}
\label{sec:beam}

At this point, and without further observations of \ulx during full orbital cycles, it is not possible to test different models such as those introduced in section \ref{sec:her}. However, we might test the extreme case where the dips are caused by partial obscuration by material very close to the Roche lobe radius, that perhaps could be at the Lagrangian points or the inner wind of the companion star. 
The Roche lobe radius, from the vantage point of the NS, will subtend an area on the sky with an angular radius $\theta$, i.e.: 
\begin{equation}
    \tan(\theta) \approx \frac{R_{\rm{RL}}}{a}{=}\frac{0.49q^{2/3}}{0.6q^{2/3} + \ln(1 + q^{1/3})},
\end{equation}
\noindent where $R_{\rm{RL}}$ is the donor radius that has filled its Roche lobe, $a$ is the separation between the donor and the NS, and their ratio depends only on the mass ratio $q{=}M_{\rm{donor}}/M_{\rm{NS}}$ \citep{1983ApJ...268..368E}. The mass ratio  in \ulx is $\gtrsim6$ \citep{2020ApJ...895...60R}. Although the mass ratio depends on the orbital inclination $i$ ($i{=}0$, refers to a face-on system), $\theta$ is actually weakly dependent on it. Assuming the companion always fills its Roche lobe, for $i$ 90$^o$-45$^o$ we find $\theta\sim28.2^o-29.4^o$. 
Given that $a\simeq2R_{\rm{RL}}$, the duration of an eclipse in a 2 day circular orbit would last $\sim$8\,h (0.16 in phase), which is comparable to the duration of the dips (see Fig.~\ref{fig:3}). 
These estimates show that the companion may indeed obscure a small opening angle from the NS vantage point of view.
Nevertheless, it is unclear if this configuration can constrain the opening angle of the funnel walls created by outflow in ULXPs (see Fig.~\ref{fig:1}). Assuming the funnel axis is fixed perpendicular to the orbital plane, its opening angle should be large enough to allow dips to be created by material in the line of sight. This would mean that the funnel's half opening angle should be $\sim$60$^o$ (full opening of $\sim$120$^o$). However, given the known super-orbital modulation in \ulx and ULXPs in general, the funnel orientation should change within the super-orbital cycle. Thus this would not exclude smaller opening angles for the funnel. Nevertheless, a large opening angle is consistent with the large truncation radius of the disk in \ulx, that was derived from its temporal properties and is consistent with the NS rotating near equilibrium \citep[e.g.][]{2020MNRAS.491.4949V,2020ApJ...899...97E}. 
Similar findings that suggest low or no beaming at all have been discussed based on spectro-temporal properties of ULXPs \citep[e.g.~NGC 300\,ULX-1][]{2018A&A...620L..12V,2019MNRAS.488.5225V}, or pulse profile evolution of X-ray pulsars during super-Eddington outbursts \citep[e.g.][]{2018A&A...614A..23K,2020MNRAS.494.5350V}.
This would mean that the proposed relation between mass accretion rate and beaming factor $b$, i.e.~$b\simeq(\dot{M}/\dot{M}_{\rm edd})^2 /73$ \citep[][]{2017MNRAS.468L..59K}, would need to be revisited in the context of ULXPs. 
Small beaming is also consistent with the detection of pulsations in ULXPs, as large beaming factors would otherwise result in very small pulsed fractions \citep{2020arXiv201109710M}. 
The above suggest that strong beaming is not needed for ULXPs, and this should be considered in the framework of ULX population synthesis \citep[e.g.][]{2020arXiv201003488K,2020A&A...642A.174M,2020ApJ...902..125A}, and perhaps gravitational wave progenitors, that are thought to go through a ULX phase \citep{2017A&A...604A..55M}. 

% and perhaps this discouraged further efforts to search for pulsations in such systems.

An important consequence of the discovery of the X-ray dips is that, regardless of assumptions about beaming, it is now evident that ULXPs can be seen even as near edge-on systems. In the literature, ULXs that show X-ray eclipses are often assumed to host BHs \citep[e.g.][]{2016ApJ...831...56U}.
Nevertheless, for another confirmed ULXP, NGC 7793 P13, it has been speculated that X-ray dips that appear during its orbital light curve were also evident of an edge-on system \citep{2014Natur.514..198M}. 
Thus, future search for pulsations should not be discouraged even for eclipsing ULXs.
Finally, another way of approaching the problem would be to search for optical dips caused by obscuration by the NS and its accretion disk \citep{2013A&A...554A...1M}.
For \ulx we can potentially confirm the low inclination via the search for optical eclipses via observations with Hubble or James Webb space telescopes \citep{2006SSRv..123..485G}.

\subsection{Implications for orbital modulation}

Our findings suggest that the mass accretion rate in \ulx is indeed super-Eddington. An implication of high accretion rates in ULXs is the change of the binary orbital period over time \citep{2020ApJ...891...44B}. According to \citet{2020ApJ...891...44B} the orbital period derivative should be: 
\begin{equation}
    \dot{P}_{\rm{orb}} \approx -3.5\times10^{-8} \left(\frac{M_{\rm{NS}}}{1.4M_{\odot}}\right)^{-1}\left(\frac{\dot{M}}{100\dot{M}_{\rm{Edd}}}\right)~s/s
\end{equation}
% \noindent 
Given that mass accretion rate for \ulx is about $30\dot{M}_{\rm{Edd}}$, the binary orbit should change by $\sim$0.3 s/y.
The binary completes $\sim$180 revolutions per year which would translate to a drift in the epoch of $T_{\rm asc}$ of the order of 30 s/year, or about 250 s between 2012 and 2020.
Future observations with X-ray telescopes could help constrain this drift by tracking the eclipses, or pulsar timing techniques \citep[e.g.][]{2020ApJ...891...44B,2020ApJ...895...60R}.

\section{Conclusion}

By analysing archival \cxo and \swift/XRT data we investigated the super-orbital and orbital variability of \ulx. 
The 2012 \cxo data obtained within 33 days show an extended low flux state, in contrast to the super-orbital clock the system.
A similar low flux state is also seen in the 2020 \swift/XRT monitoring data.
These off-states might be related to propeller transition similar to ULXP NGC 5907 ULX1.
Alternatively they could be indicative of a variable super-orbital period like those in other accreting pulsars (see Her~X-1, SMC~X-1).
Moreover, we have reported the presence of periodic dips in the \cxo X-ray light curve of \ulx. 
Although X-ray dips are also seen in bright X-ray binaries \citep{2017ApJ...851L..27M} and ULXs \citep{2018MNRAS.477.3623W}, this is the first evidence of such property in ULXPs.
The physical origin of the dips remain unclear, however they could be related to a plethora of mechanisms that have been proposed to explain similar features in HMXBs. Our finding demonstrates the need for developing numerical simulations of HMXB systems in the context of super-Eddington accretion and investigating these intriguing phenomena. From an observational point of view, it demonstrates the need for long monitoring observations of ULXPs and ULXs to identify and confirm the presence of features related to orbital modulation. Such combined efforts would help to develop a physically motivated, self-consistent model able to explore the central engines of ULXPs.

\section*{Acknowledgements}
The authors would like to thank the anonymous referee for their comments and suggestions that helped improve the manuscript.
We would like to thank the organizers of the ``Chandra Frontiers in Time-Domain Science'' meeting that was held virtually in October 2020. Presentation of M51 monitoring data and discussions that followed resulted in the current work.\\
GV acknowledges support by NASA Grant numbers 80NSSC20K1107, 80NSSC20K0803 and 80NSSC21K0213.\\
Software used: HEASoft v6.26, CIAO v4.12.1, Python v3.7.3, IDL\textsuperscript{\textregistered}

\bibliography{general.bib}
\bibliographystyle{aasjournal}



\end{document}